\documentstyle[epsfig, rotate]{elsart} 
\begin{document}
\def\figuras{.}
\def\altura{4.6cm}
\begin{frontmatter}
\title{Ensemble equivalence for non-extensive thermostatistics} 
\author{Ra\'ul Toral$^{\ref{r1}}$, Rafael Salazar$^{\ref{r2}}$}   
\address[r1]{ 
Instituto Mediterr\'aneo de Estudios Avanzados (IMEDEA), 
UIB--CSIC, Campus UIB, 07071-Palma de Mallorca, Spain.} 
\address[r2]{ 
Schuit Institute of Catalysis (SKA), Eindhoven University of
Technology, P.O. Box 513, 5600 MB Eindhoven, The Netherlands.}

\begin{abstract}  
We compute the internal energy of different Ising type models, both long--range and short--range, under Tsallis statistics using the microcanonical and the canonical ensembles and we discuss under which conditions both ensembles give equivalent results. 
\end{abstract}
\vspace{-0.5cm}

\end{frontmatter}

\vspace{-1.0cm}
\section{Introduction} \label{s1} 
\vspace{-0.5cm}

In this paper we study different Ising type
systems in the microcanonical and the canonical ensembles under the generalized
statistics introduced by Tsallis \cite{tsa88}. This is based upon the following
alternative expression for the entropy: 

\begin{equation}\label{e1} 
S_q=\frac{1- \sum_{i} p_i^q}{q-1},  
\end{equation}  
\vspace{-0.5cm}

depending on a parameter
$q$. A probability $p_i$ is assigned to each of the $i=1, \dots, W$ microscopic
configurations. The set of probabilities $\{p_i\}$ must be obtained maximizing $S_q$ under the appropriate
constraints. The physical observables associated to a microscopic dynamical
function ${\cal O}$ are computed as statistical averages $\langle {\cal O}
\rangle$ obtained using the rule\cite{tsa98}: $
\langle {\cal O}\rangle = \sum_{i} O_i P_i~$ with $P_i = {p_i^q}/{\sum_{j} p_j^q}$ and $O_i$ is
the value of ${\cal O}$ at the configuration $i$ whose probability is $p_i$.
Different statistical ensembles give rise to different probabilities $p_i$ and
we address in this paper the question of equivalence between the microcanonical
(fixed energy) and the canonical (fixed average energy) ensembles. To be more
specific we consider Ising type systems with Hamiltonian:  
\clearpage

\begin{equation}\label{e14}  
{\cal H} = \sum_{(n,m)} \frac{1-s_n s_m}{r_{nm}^{\alpha}}.  
\end{equation}  
\vspace{-0.5cm}

Where $s_n=\pm 1$, and the indexes
$n,m=1,\dots,N=L^d$ run over the distinct pairs of sites of a regular $d$-dimensional lattice of lattice spacing equal to $1$ and with periodic boundary conditions;
$r_{nm}$ is the  distance between the spins $n$ and $m$, and the parameter
$\alpha$ sets the interaction range. The energy of configuration $i$ is 
$\varepsilon_i$ (notice that the ground state has zero energy). The usual short
range Ising model is recovered in the limit $\alpha \to \infty$ where
$1/r_{nm}^{\alpha}\to 0$ unless $n$ and $m$ are nearest neighbors with
$r_{nm}=1$. For $\alpha \le d$ the average energy per particle diverges in the
thermodynamic limit and the system is said to be non--extensive. More
precisely, a convenient scale for the average energy per spin in a finite
system of size $N$ is given by \cite{jun95}: $E/N\sim \tilde N=
\frac{N^{1-\alpha/d}-\alpha/d} {1-\alpha/d}$. We see that for $\alpha>d$ the
average energy per spin scales as a constant in the limit of large $N$, whereas
for $\alpha \le d$, it grows with the system size, a non--extensive behavior. 
Throughout this paper we will be considering the following cases: 

(a) $\alpha=\infty$, $d=1$, the short range one dimensional Ising model. \\ 
(b) $\alpha=\infty$, $d=2$, the short range two dimensional Ising model. \\ 
(c) $\alpha=0.8$, $d=1$, a non extensive, long range, one dimensional Ising 
model. \\ 
(d) $\alpha=0$, the non extensive infinite range Ising model, whose
properties are basically independent of the spatial dimension $d$. 

For each of these cases we will compute the internal energy $E(T,N)$ (the
average value of the Hamiltonian $\langle {\cal H}\rangle$) as a function of
the temperature $T$ and the number of particles $N$, in the microcanonical
and the canonical ensembles. We will compare the results obtained in both
ensembles using the standard definition of temperature as well as a recent
proposal for a {\sl physical} temperature \cite{abe01}.

\vspace{-0.8cm}
\section{The microcanonical ensemble} 
\vspace{-0.5cm}

The microcanonical ensemble is defined by
fixing the energy $E$ and setting \makebox{$p_i=0$} to those configurations
whose energy is not equal to $E$. The maximization problem for the original
entropic form $S_q$  given by Eq.(\ref{e1}) with the constraint of given energy
$E$, and the normalization condition $\sum_i p_i = 1$, has the solution of 
equiprobability: 

\begin{equation}
p_i=\left\{\begin{array}{ll}\Omega(E,N)^{-1}, &  ~~~~\varepsilon_i = E, \\ 
0, & ~~~~{\rm otherwise.} \end{array}\right. 
\end{equation} 
\vspace{-0.5cm}

$\Omega(E,N)$ is the number of configurations with energy $E$ for a system with
$N$ particles. Using this solution, the entropy as a function of the energy is
given by$ S_q(E,N)= (\Omega(E,N)^{1-q}-1)/(1-q)$. Finally, the temperature is
defined by the thermodynamic relation $\frac 1 T= \frac {\partial S_q}{\partial
E}= \Omega(E,N)^{-q}\frac{\partial \Omega(E,N)}{\partial E}$.
This relation allows us to plot the energy as a function of the
temperature, $E(T,N)$. 

We see that an essential ingredient in the microcanonical ensemble is
$\Omega(E,N)$, the number of configurations having energy $E$ for a system with
$N$ particles. The exact form of this function depends on the interaction range
parameter $\alpha$ and on the spatial dimension $d$, although the total number
of configurations is $\sum_E \Omega(E,N)=2^N$ independent of $\alpha$ and $d$. 
We now specify this function in each of the four cases of interest: \\
(a) $\alpha=\infty$, $d=1$. The possible energy levels are $E_k=4k$,
$k=0,\dots,N/2$ (assuming that $N$ is an even number) and their degeneracy is easily computed as: $\Omega(E_k,N)=2 {N \choose 2k}$.\\
(b) $\alpha=\infty$, $d=2$. The possible energy values are $E_k=4k$, 
$k=0,\dots,N$.
No closed exact expression is known for the function $\Omega(E,N)$. However, the
exact solution for the partition function of the $L\times M$ Ising model
\cite{kaufman}, has allowed Beale \cite{bea96} to write a Mathematica program
that can actually compute $\Omega(E,N)$ for moderate values of $N$. Using this
program we have generated the exact values of $\Omega(E,N)$ for $N=32^2$ or
smaller.\\ 
(c) $\alpha=0.8$, $d=1$. This is the more complicated case,
because no analytical expression or exact numerical values is available. For small values of $N$, up to
$N=34$ we have made a complete enumeration of the $W=2^N$ configurations and their energies $\varepsilon_i$. For larger sizes, up to $N=3000$, the values of
$\Omega(E,N)$ have been obtained by using a numerical sampling method known as
Histogram by Overlapping Windows (HOW)\cite{bha87}. Details of the implementation of the method for this particular problem can be found in \cite{sal01}.\\ 
(d) $\alpha=0$. This case is equivalent to consider the Bragg--Williams
approximation to the solution of the Ising model \cite{chaikin}. The energy
levels are $E_k=2k(N-k)$, $k=0,\dots,N/2$ (assuming again that $N$ is an even
number) and the number of states is: $\Omega(E_k,N)=  2{N \choose k}$ for $ k=0,1,\dots, N/2-1$ and $\Omega(E_k,N)={N\choose N/2}$ for $k=N/2$.

\vspace{-0.5cm}
\section{The canonical ensemble}
\vspace{-0.5cm}

In the canonical ensemble we fix a value for the average energy $E=\langle {\cal
H}\rangle$. Maximization of the entropy, Eq.(\ref{e1}), under this 
constraint and the normalization condition $\sum_i p_i = 1$, leads to
the following solution for the probabilities $P_i$ \cite{tsa98}:  

\vspace{-0.1cm}
\begin{equation}\label{e9} 
P_i = \left \{ \begin{array}{ll} 
\frac {[1-(1-q)
\varepsilon_i/T']^{\frac q {1-q}}}   
{\sum_{j} [1-(1-q)\varepsilon_j/T']^{\frac q {1-q}}}, 
& \hspace{1.0cm}1- (1-q)\varepsilon_i/T' >  0, \\
0, &\hspace{1.0cm}{\rm otherwise.}\end{array} \right. 
\end{equation} 
\vspace{-0.3cm}

Where we have defined 
\clearpage

\begin{equation}\label{e10}  
T'= (1-q)\sum_{j} \varepsilon_j P_j + T(\sum_{j} P_j^{1/q})^{-q}  
\end{equation} 
\vspace{-0.5cm}

and $T$ is the
(inverse of the) Lagrange multiplier used to enforce the condition of fixed
average energy in the maximization procedure. Finally, it is possible to show the validity of the the Legendre
structure of the resulting thermodynamics formalism by proving the relation \cite{pla97b,tsa98}: $1/T=\partial S_q/\partial E$.

Although Eqs.(\ref{e9},\ref{e10}) form a closed set, it is very difficult to use them in that form because the number of terms in the sums is the number of microscopic configurations, or $2^N$, an extremely large number. It is possible to simplify the calculations by rewriting those equations in terms of the number of states $\Omega(E,N)$. More precisely, one can replace the sum over microscopic configurations (with $2^N$ terms) by sum over energy levels (with an number of terms proportional to $N\tilde N$),  $\sum_j\longrightarrow \sum_k\Omega(E_k,N)$.
Using the function $\Omega(E,N)$ it is then possible to compute numerically the different sums and hence to compute the internal energy (and other quantities of interest) within the canonical ensemble formalism\cite{sal01}.

\vspace{-0.5cm}
\section{Results}
\vspace{-0.5cm}

\begin{figure}[!ht]  
\makebox{\hspace{-1.0truecm}{\epsfig{figure=\figuras/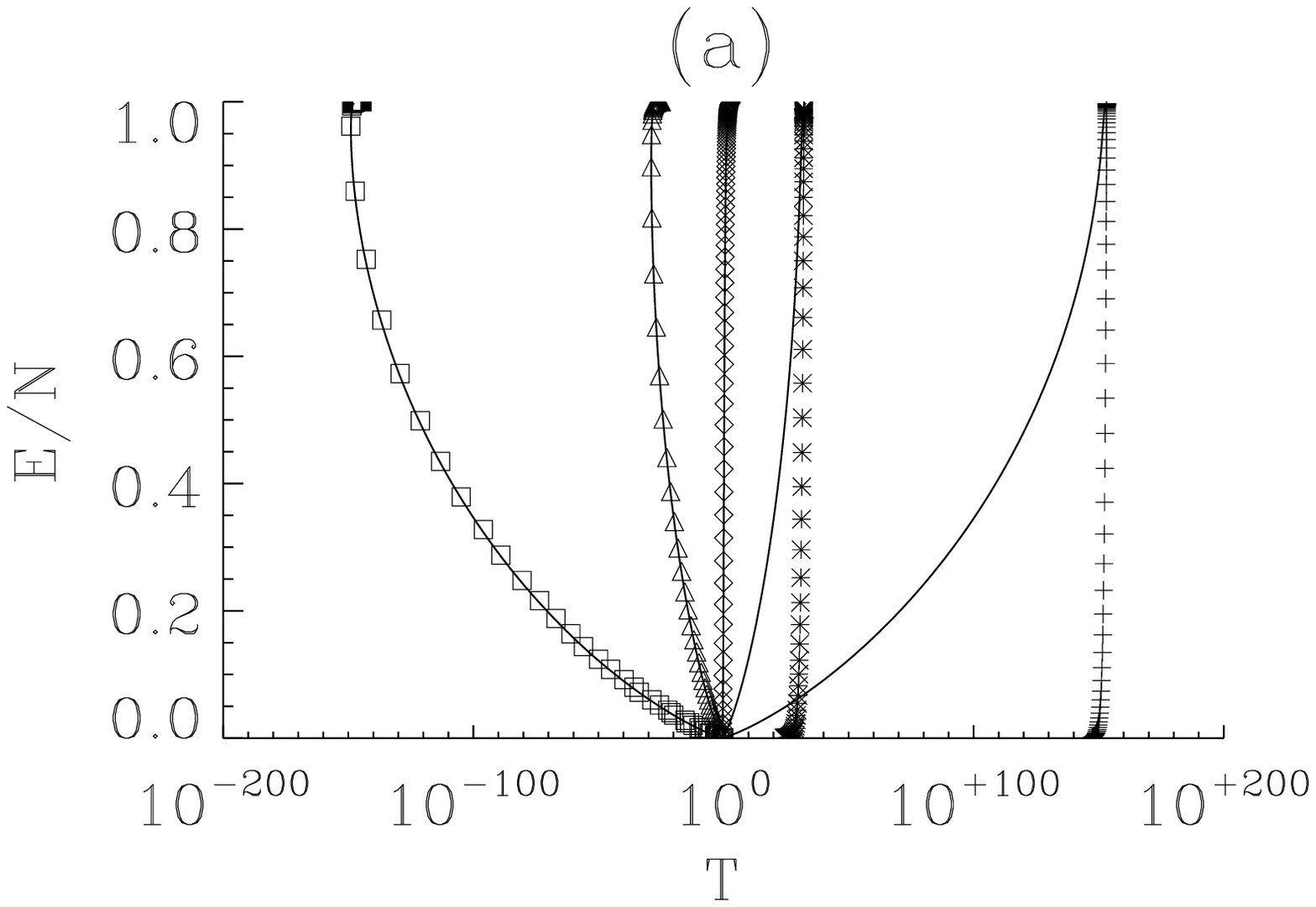,width=8cm,height=\altura}}
\hspace{-1.0truecm}{\epsfig{figure=\figuras/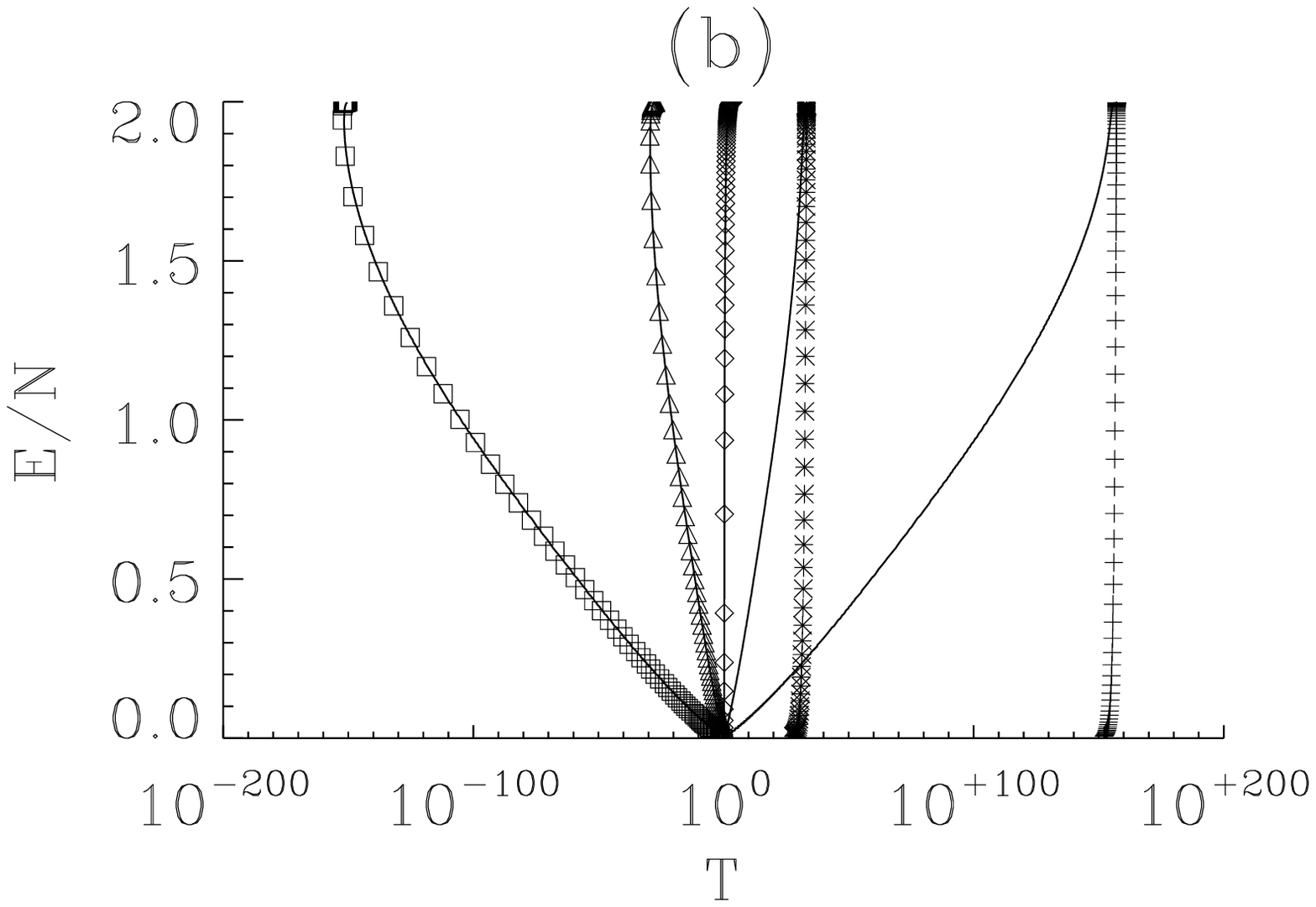,width=8cm,height=\altura}}}
\makebox{\hspace{-1.0truecm}{\epsfig{figure=\figuras/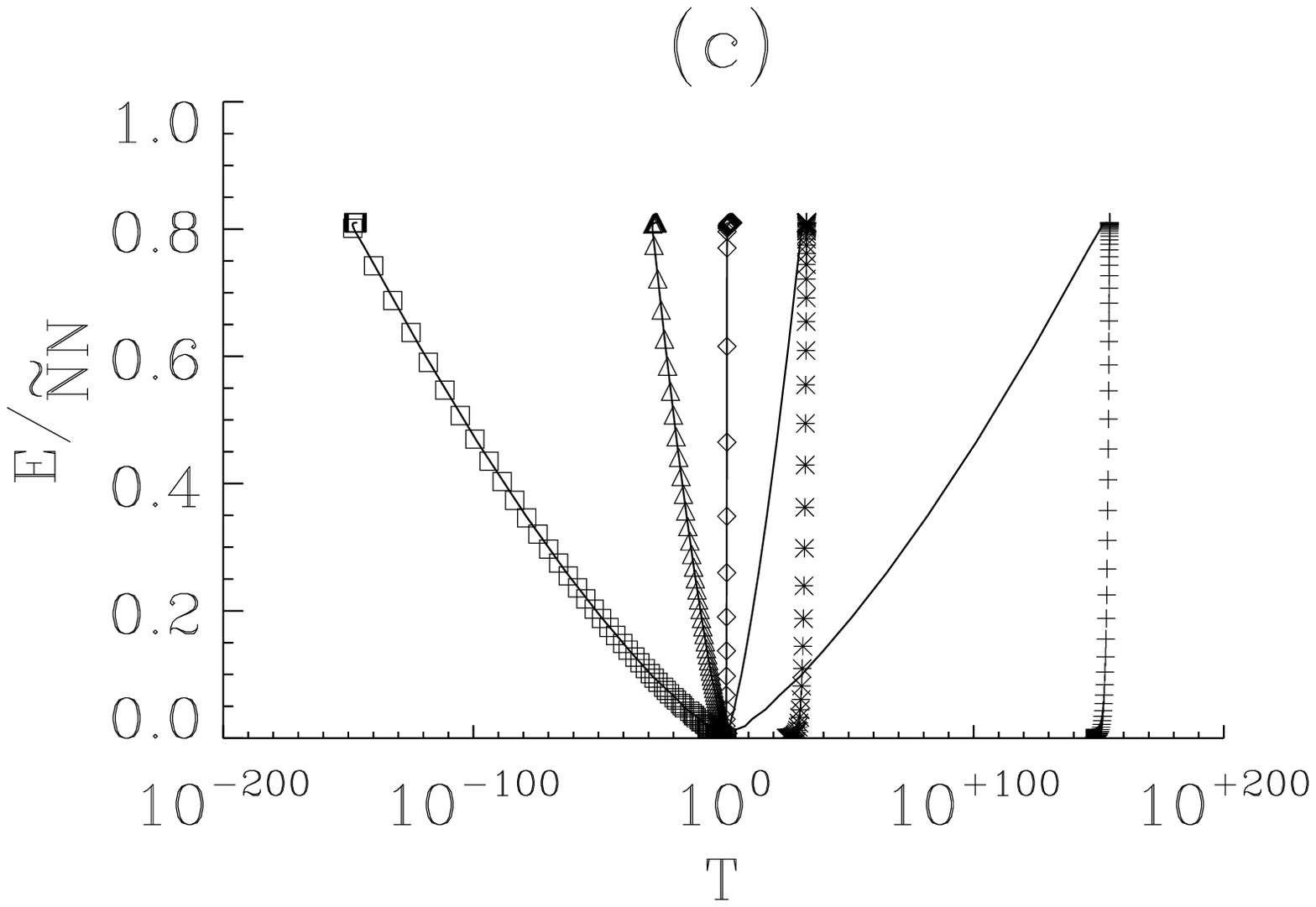,width=8cm,height=\altura}}
\hspace{-1.0truecm}{\epsfig{figure=\figuras/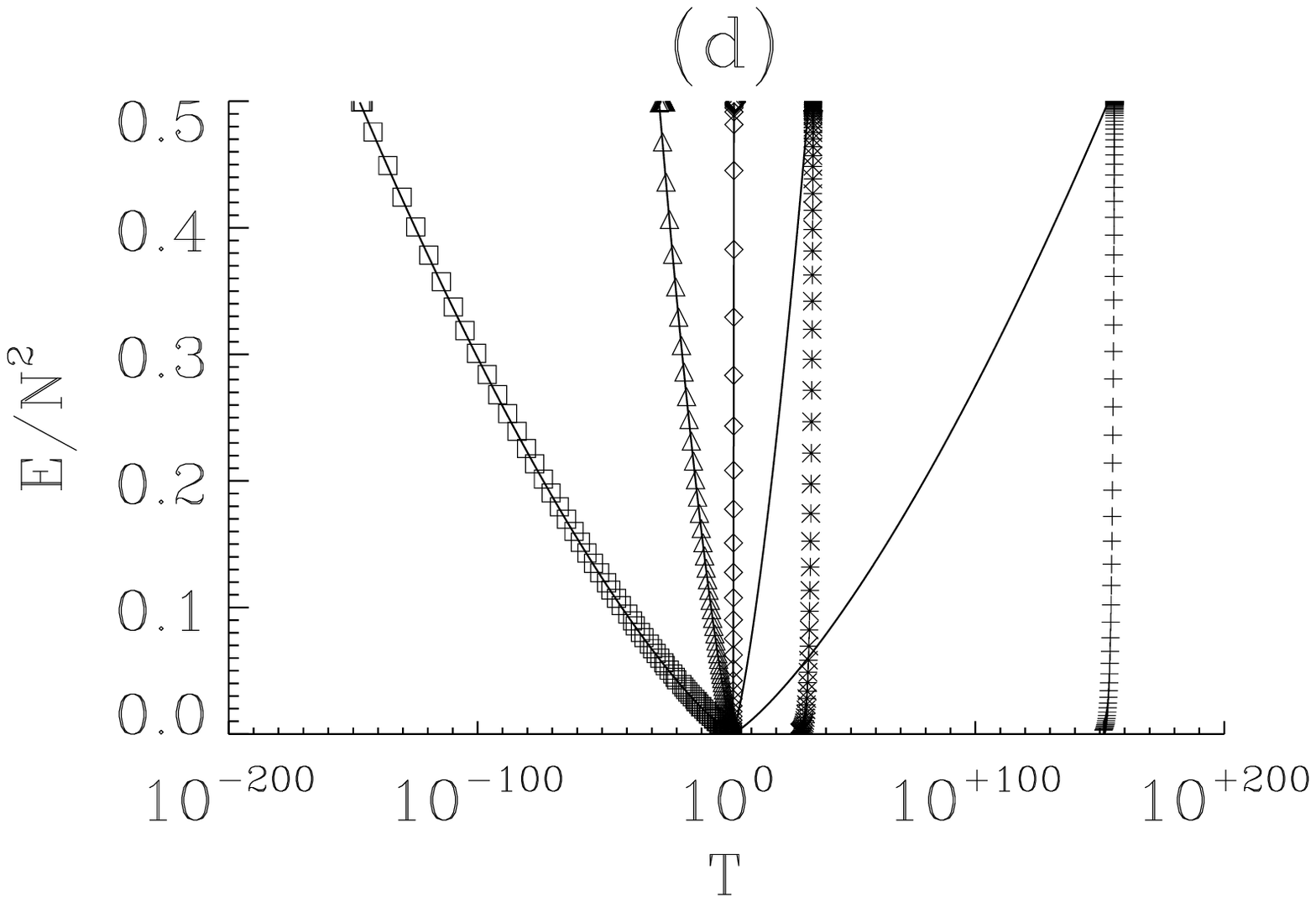,width=8cm,height=\altura}}}
\caption{
Internal energy $E$ as a function of the temperature $T$ for the Ising model defined by (\ref{e14}) in the cases 
(a) $\alpha=\infty$, $d=1$,
(b) $\alpha=\infty$, $d=2$,
(c) $\alpha=0.8$, $d=1$,
(d) $\alpha=0$. The number of spins is $N=1000$. Solid lines are the results obtained in the microcanonical ensemble, while symbols correspond to the canonical ensemble.
From left to right curves, the values of $q$ are
$q=0.5,0.9,1,1.1,1.5$.
\label{f1}} 
\end{figure}

In Fig.\ref{f1} we plot the internal energy versus the
temperature, in the microcanonical and the canonical ensembles obtained using
the appropriate number of states $\Omega(E,N)$ for each case (a-d).
It is shown that, within the accuracy displayed in that figure and in all cases, the microcanonical and canonical results agree for $q\le 1$ but disagree for $q>1$. 

\begin{figure}[!ht]  
\makebox{\hspace{-1.0truecm}{\epsfig{figure=\figuras/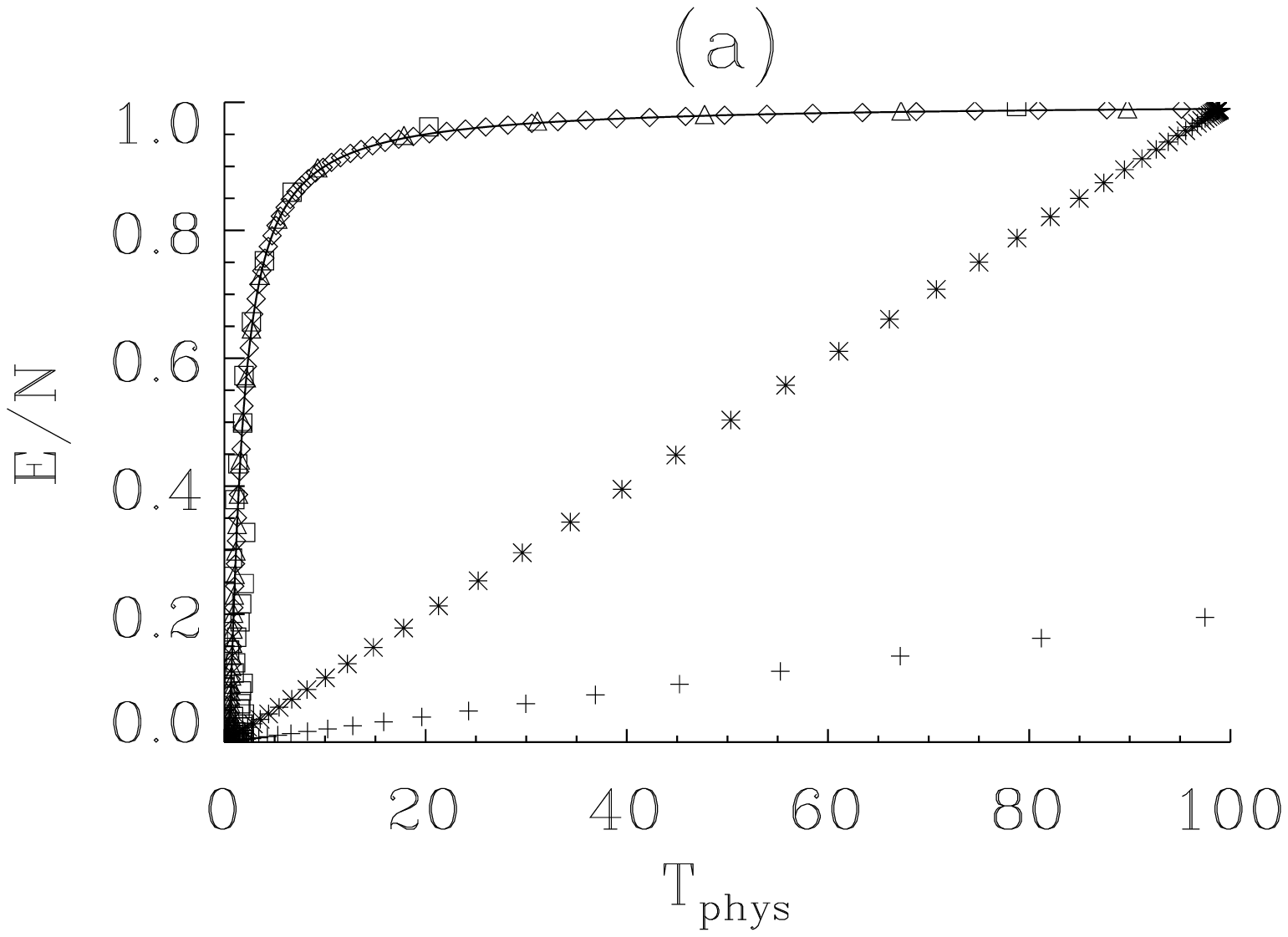,width=8cm,height=\altura}}
\hspace{-1.0truecm}{\epsfig{figure=\figuras/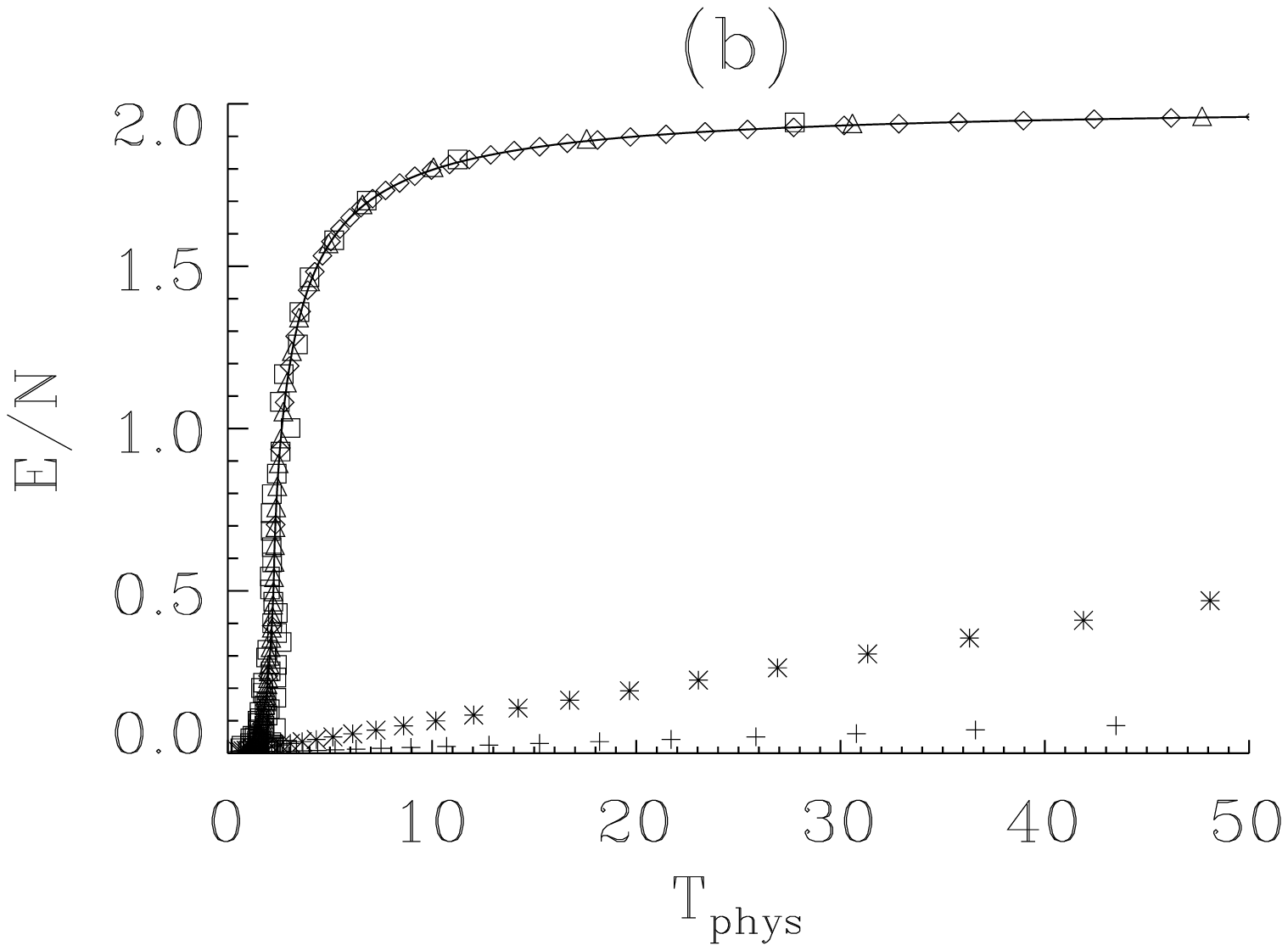,width=8cm,height=\altura}}}
\makebox{\hspace{-1.0truecm}{\epsfig{figure=\figuras/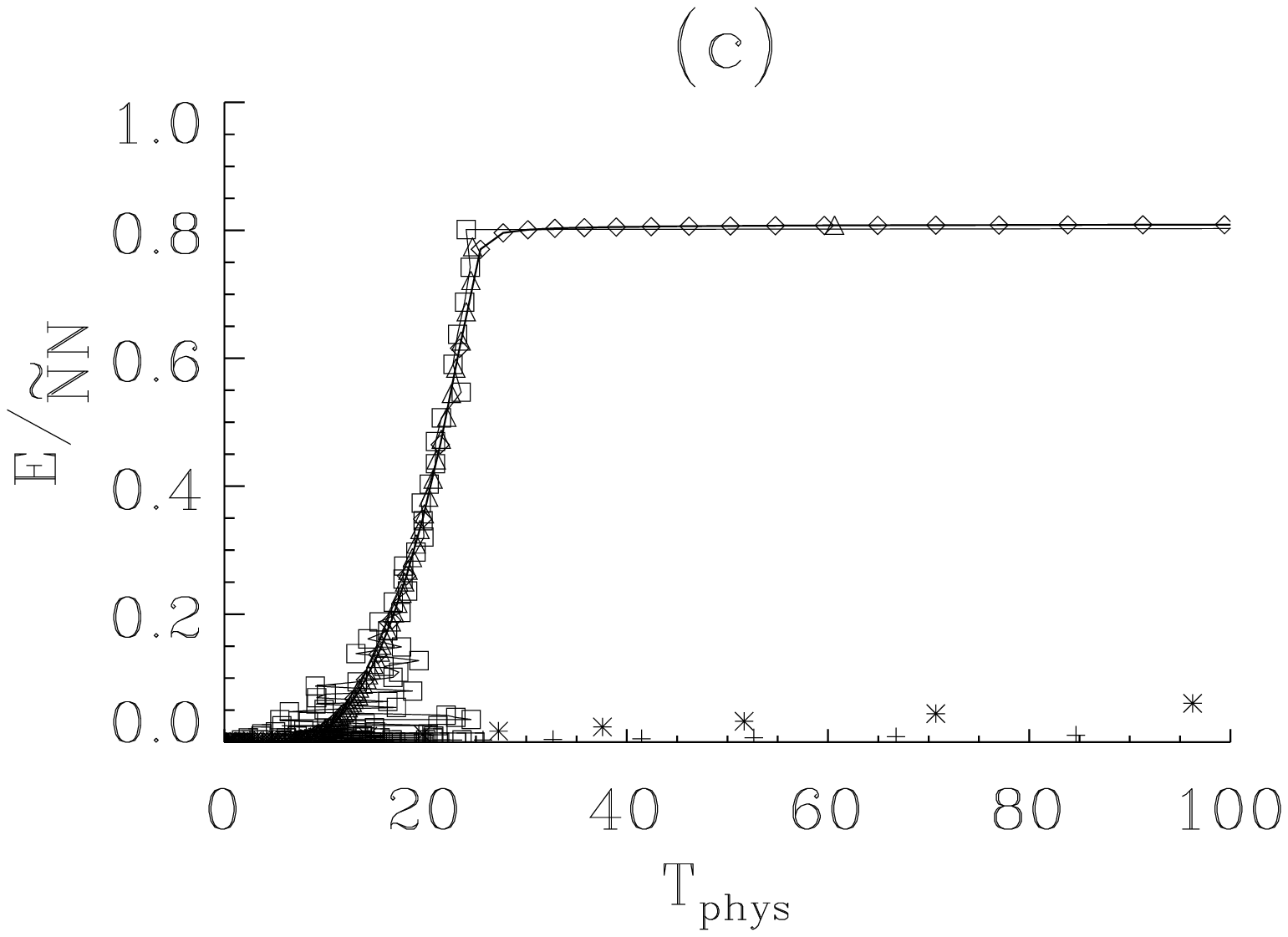,width=8cm,height=\altura}}
\hspace{-1.0truecm}{\epsfig{figure=\figuras/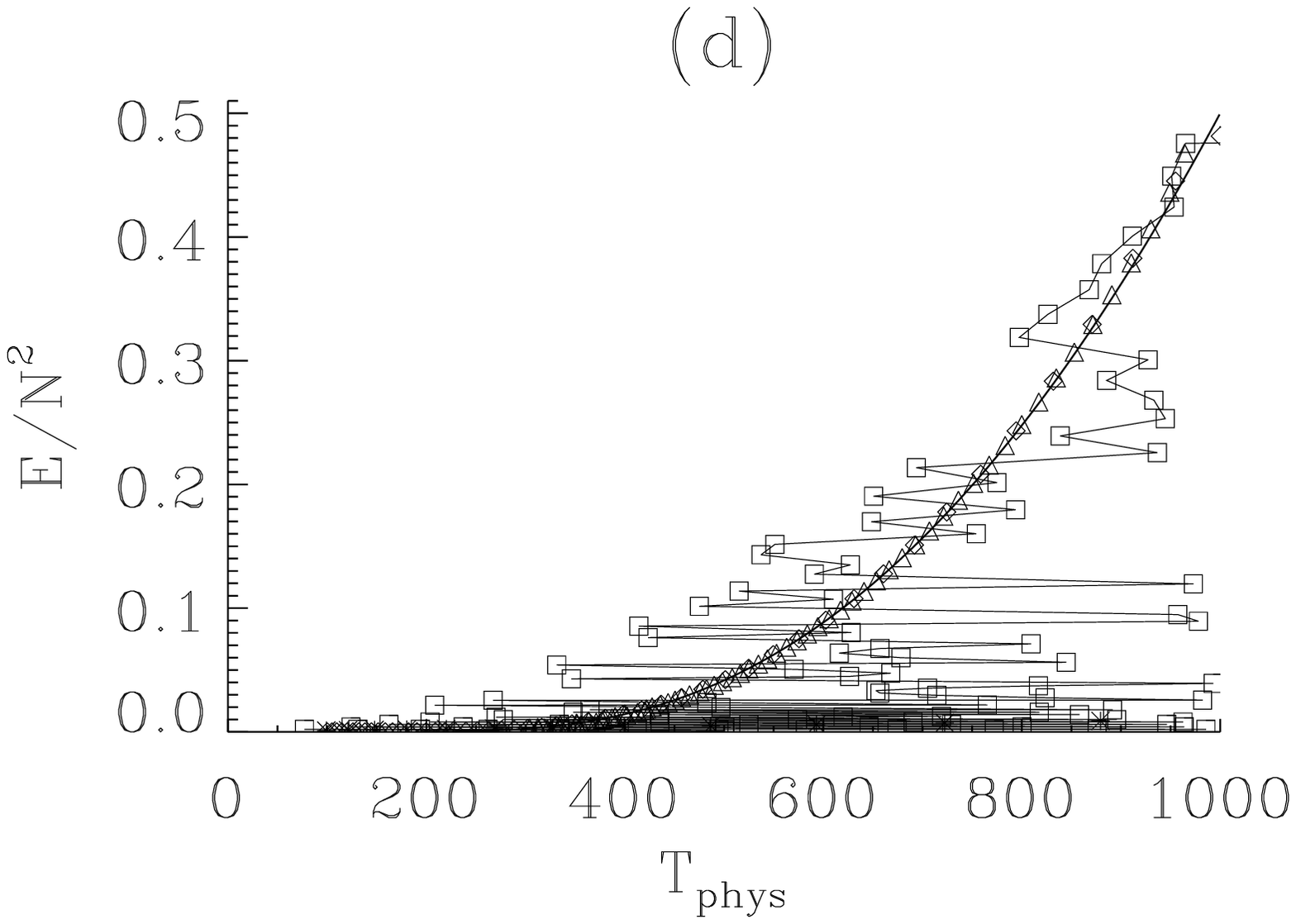,width=8cm,height=\altura}}}
\caption{
Internal energy $E$ as a function of the physical temperature $T_{phys}$.
Same cases and symbol meanings as in Fig.\ref{f1}.
\label{f2}}
\end{figure} 

Recently \cite{abe01} it has been suggested that an appropriate definition for a physical temperature is

\vspace{-0.1cm}
\begin{equation}\label{tphys}
T_{phys}=\left(1+\frac{1-q}{k}S_q\right)\left(\frac{\partial S_q}{\partial U_q}\right)^{-1}.
\end{equation}
\vspace{-0.3cm}

In Fig. \ref{f2} we plot the internal energy as a function of this physical temperature in the four cases of interest (a-d). We can see from this figure that the results of the microcanonical ensemble collapse in a single curve for all values of $q$, i.e. the dependence on $q$ disappears when using the physical temperature in the microcanonical ensemble. This collapse is in agreement with a recent theoretical analysis of the definition of physical temperature\cite{tor01}. The results of the canonical ensemble show some zig-zags which should be repaired by using a Maxwell--type construction (although we have not been able to find an unambiguous way of performing such a construction). Those zig-zags have their origin in the necessary transformation to go from the parameter $T'$ to the temperature $T$ as given by Eq. (\ref{e10}), and a plot of $T$ versus $T'$ already shows that behavior. The mathematical origin of the zig-zags lies in the condition $1-(1-q)\epsilon_i/T' > 0$ of Eq. \ref{e9}: when we sweep along the values of $T'$, the number of different energy levels with a non--zero probability increases in a discrete way, thus producing steps or zig-zags.
The number of these steps is proportional to $N \tilde N$, and in the thermodynamic limit we will get an almost continuous set of energy levels.
These zig-zags are more pronounced the larger the interaction range. Except for this zig-zag behavior, we see again that the microcanonical and the canonical ensemble results agree for $q\le 1$ and disagree for $q >1$.
The same conclusions are reached when studying the system magnetization instead of the internal energy. A more detailed study, beyond the scope of this paper, should consider also other properties such the heat capacity or the magnetic susceptibility.

Notice that the ultimate reason for not having equivalence between the two ensembles for $q>1$, is
that fluctuations of the energy in the canonical ensemble can not be neglected.
We have checked that this is indeed the case by computing the energy
fluctuations $\sigma({\cal H)}=\sqrt{\langle {\cal H}^2\rangle -\langle {\cal
H}\rangle^2}$ as a function of the system size. We have checked that the fluctuations, when normalized by the scale of energy, $N\tilde N$, do not decay
to zero for increasing $N$ in the range of temperatures for which the
microcanonical and canonical ensemble do not agree.  For $q \le 1$ fluctuations
do decay to zero with the system size in all the temperature range.

We acknowledge financial support from DGES (Spain) project PB97-0141-C02-01 and
MCyT (Spain) project BMF2000-0624.

\end{document}